\documentclass[preprint2,twocolappendix,trackchanges]{aastex6}

\usepackage{booktabs}
\usepackage{color}
\usepackage{soul}
\usepackage{ulem}
\usepackage{bm}
\usepackage{amsmath}

\begin{document}


\title{Energy Levels, Lifetimes and Transition rates for P-like ions from  C\lowercase{r}~X to Z\lowercase{n}~XVI from large-scale Relativistic Multiconfiguration Calculations}

\author{K. Wang\altaffilmark{1,2,3}, P. J\"onsson\altaffilmark{1}, G. Gaigalas\altaffilmark{4}, L. Rad{\v{z}}i{\={u}}t{\.{e}}\altaffilmark{4}, P. Rynkun\altaffilmark{4}, G. Del Zanna\altaffilmark{5}, C. Y. Chen\altaffilmark{3}}

\affil{	$^1$Group for Materials Science and Applied Mathematics, Malm\"o University, SE-20506, Malm\"o, Sweden;  {\color{blue} per.jonsson@mah.se}\\
	$^2$Hebei Key Lab of Optic-electronic Information and Materials, The College of Physics Science and Technology, Hebei University, Baoding 071002, China\\
	$^3$Shanghai EBIT Lab, Institute of Modern Physics, Department of Nuclear Science and Technology, Fudan University, Shanghai 200433, China; {\color{blue} chychen@fudan.edu.cn}\\
	$^4$Institute of Theoretical Physics and Astronomy, Vilnius University, Saul\.etekio av. 3, LT-10222, Vilnius, Lithuania\\
    $^5$DAMTP, Centre for Mathematical Sciences, University of Cambridge, Wilberforce Road, Cambridge CB3 0WA, UK}




\begin{abstract}
The fully relativistic multiconfiguration Dirac--Hartree--Fock method is used to compute excitation energies and lifetimes for the 143 lowest states of the 
$3s^23p^3$, $3s3p^4$, $3s^23p^23d$, $3s3p^33d$, $3p^5$, $3s^23p3d^2$ configurations in P-like ions from \ion{Cr}{10} to \ion{Zn}{16}.  
Multipole (E1, M1, E2, M2) transition rates, line strengths, oscillator strengths, and branching fractions among these states 
are also given. Valence-valence and core-valence electron correlation effects are systematically accounted for using large basis function expansions.  
Computed excitation energies are compared with the NIST ASD and CHIANTI compiled values and previous calculations. The mean average absolute difference, removing obvious outliers, 
between computed and observed energies for the 41 lowest identified levels in Fe XII is only \mbox{0.057 \%}, implying 
that the computed energies are accurate enough to aid identification of new emission lines from the sun and other astrophysical sources. 
The amount of energy and transition data of high accuracy is significantly increased for several P-like ions of astrophysics interest, where experimental data are still very scarce.  
\end{abstract}

\keywords{atomic data - atomic processes}


\section{INTRODUCTION}\label{sect:in}
P-like ions of the iron group elements have prominent lines in the ultraviolet (UV) and
extreme ultraviolet (EUV) spectral regions that are used for plasma diagnostics, especially 
to measure electron densities. 
\mbox{Fe XII} is especially important for the solar corona, as it produces the most prominent lines
in the EUV, as observed by e.g. Skylab, SOHO,  and more recently with the 
EUV Imaging Spectrometer (EIS) instrument on board the Hinode satellite \citep[see, e.g.][]{YHINODE,DZHINODE}.
\mbox{Fe XII} lines have also been observed by  Chandra and XMM-Newton \citep{ChandraXMM}. 

Because of their complex atomic structure, the radiative and scattering 
(by electron impact) calculations of P-like ions are notoriously difficult. 
Because of its importance, much effort has been produced  to provide 
accurate scattering data for Fe XII, see for example \cite{SS2005,DZFEXII}.
It is only with the most recent calculations that large discrepancies between 
predicted and observed line emission has largely been resolved. 
Only about half of the $3s^23p^23d$ levels in Fe XII were identified, mostly by B.C.Fawcett and 
collaborators \citep[cf.][]{bromage_etal:78} in the 1970's, from the 
strongest decays in the EUV observed in theta-pinch spectra. 
\cite{DZ05} reviewed the identifications and suggested several new ones,
using the \cite{SS2005} atomic data. 

Important information for some of the states of the $3s^23p^23d$ configuration 
has been obtained from  fast-beam spectroscopy. For example, 
radiative lifetimes  in Fe XII, Co XIII and Cu XV were measured by \cite{LIFE}. 
For a recent review on Fe XII see \cite{Traebert.2008.V130.p12018}.
As noted by \cite{Vilkas}, there are however some differences between calculated and measured lifetimes, 
calling for a reexamination of the latter. 

Much effort has also been devoted to the calculation of radiative data
for  Fe XII, also to aid the identifications  \citep[see, e.g.][]{FROESEFISCHER2006607,Vilkas,Tayal.2011.V97.p481}.
The identifications and atomic data for the Fe XII lines have recently been reviewed 
by \cite{BD2014} using EBIT spectroscopic measurements, noting one significant 
discrepancy with the identifications proposed by \cite{DZ05}.

For the other P-like ions, very few atomic data are available, with the exception of 
a few low charge states and   Ni XIV. 
Although less studied, also Ni XIV has diagnostic value and atomic data 
- electron impact collision strengths, energy levels, oscillator strengths, and spontaneous radiative decay
rates - have been calculated by \cite{NiXIV}.

The purpose of the present work is to provide transition energies for the lowest 143 states in P-like ions with spectroscopic accuracy, 
i.e. the accuracy is high enough so that the energies can be directly used to identify  lines from laboratory or space observations.
 Further, the present work aims at providing a consistent and accurate set of transition rates for all the ions, for modeling purposes. 
This  work is an extension of our previous efforts~\citep{Chen.2018.V206.p213,Chen.2017.V113.p258,Guo.2016.V93.p12513,Guo.2015.V48.p144020,Jonsson.2014.V100.p1,Jonsson.2013.V559.p100,Jonsson.2011.V97.p648,Wang.2018.V234.p40,Wang.2018.V208.p134,Wang.2017.V194.p108,Wang.2017.V119.p189301,Wang.2017.V187.p375,Wang.2017.V229.p37,Wang.2016.V223.p3,Wang.2016.V226.p14,Wang.2015.V218.p16,Wang.2014.V215.p26,Si.2017.V189.p249,Si.2016.V227.p16,Zhao.2017.V119.p313} to supply highly accurate atomic data for L- and M-shell systems, 
see \cite{Jonsson.2017.V5.p16} for a review.

\section{Theory and Calculations}
\subsection{MCDHF}\label{Sec:MCDHF}
In the multiconfiguration Dirac-Hartree-Fock (MCDHF) method the wave function $\Psi (\gamma PJM)$ for a state labeled $\gamma PJM$, where $\gamma$ is the orbital occupancy and angular coupling tree quantum numbers, $P$ the parity, $J$ the total angular momentum quantum number, and $M$ the total magnetic quantum number, is written as a linear combination of 
$N$ configuration state functions $\Phi(\gamma_rPJM)$ (CSFs)
\begin{equation}
	\Psi(\gamma PJM) =\sum^N_{r=1}c_r \Phi(\gamma_r P J M).
\end{equation}
The CSFs are antisymmetrized and $jj$-coupled many electron functions built from products of one-electron Dirac orbitals~\citep{Grant.2007.V.p,review_MCHF_MCDHF}.  The radial parts of the Dirac orbitals and the expansion coefficients for the targeted states are obtained by solving the MCDHF equations, which result from applying the stationary condition on the
state averaged Dirac-Coulomb energy functional with added terms to enforce orthonormality of the radial orbitals. The Breit interaction and leading QED effects (vacuum polarization and self-energy) are included in subsequent configuration interaction (CI) calculations.
\subsection{Transition parameters}\label{Sec:trans} 
Transition parameters such as transition rates $A$, line strengths $S$, and weighted oscillator strengths $gf$ between an initial
$\gamma PJM$ and a final $\gamma' P'J'M'$ state are expressed in terms of reduced matrix elements
\begin{equation}
\langle \Psi(\gamma PJ) \| {\bf T} \| \Psi(\gamma' P'J') \rangle = \sum_{r,s} c_r c'_s \langle \Phi(\gamma_r PJ) \| {\bf T} \| \Phi(\gamma'_s P'J') \rangle,
\end{equation}
where ${\bf T}$ is the transition operator. The evaluation of the matrix elements 
between separately optimized initial and final states, i.e. states that are built from different and mutually non-orthonormal orbital sets,
follows the prescription given in \cite{Olsen.1995.V52.p4499}.

For electric multipole transitions E1, E2 etc. there are two forms of the
transition operator; the length (Babushkin) form and the velocity (Coulomb) form~\citep{GRANTGAUGE}.
The length form is usually preferred. 
Following \cite{FroeseFischer.2009.V134.p14019} and \cite{atoms2020215} we use
the relative difference 
\begin{equation}
dT =  \frac{|A_l - A_v|}{\max(A_l,A_v)}
\end{equation}
of the transition rates $A_l$  and $A_v$ computed in length and velocity
form as an indicator of accuracy of the rate. It should be emphasized that the values of $dT$ do
not represent an uncertainty estimate of the rate for each individual transition.
Instead, they should be considered statistical indicators of
uncertainties within given sets of transitions. 
\subsection{Calculations}\label{Sec:GENERATE}
Calculations were performed for the 143 lowest states belonging to the
$3s^23p^3$, $3p^5$, $3s3p^33d$ odd and
$3s3p^4$, $3s^23p^23d$, $3s3p^23d^2$ even configurations. The odd and even 
states were determined in separate calculations. As a starting
point, two MCDHF calculations were performed for, respectively, the statistically weighted average of the odd and
even parity states. To include electron correlation, and improve
on the computed energies and wave functions, the initial calculations
were followed by separate MCDHF calculations for
the odd and even parity states, with CSF expansions obtained by allowing single and
double (SD) substitutions from the reference configurations to
active orbital sets with principal quantum numbers up to $n = 6$
and with orbital angular momenta up to $l = 5$ ($h$-orbitals). Only CSFs that
have non-zero matrix elements with the CSFs belonging to the
reference configurations were retained. No substitutions were
allowed from the $1s$ shell, which defines an inactive closed core.
Furthermore, the substitutions were restricted in such a way that
only one substitution was allowed from the $2s$ and $2p$ subshells
of the configurations in the MR, and thus the generated expansions
account for valence and core-valence electron correlation \citep{Sturesson.2007.V177.p539,review_MCHF_MCDHF}. The wave functions were
further improved by augmenting the above expansions with CSFs obtained by SD substitutions 
from the valence subshells of the reference configurations to active orbital sets with 
principal quantum numbers extended to $n = 8$ and with orbital angular momenta up to $l = 6$.
The neglected core-core correlation is comparatively unimportant for
both the energy separations and the transition probabilities \citep{Gustafsson.2017.V5.p3}.

The MCDHF calculations were followed by CI calculations including
the Breit-interaction and leading QED effects. The number
of CSFs in the final odd and even state expansions were
approximately 3~250~000 and 4 600 000, respectively, distributed
over the different $J$ symmetries.
All calculations were performed with the GRASP2K code
\citep{Jonsson.2007.V177.p597,Jonsson.2013.V184.p2197}. To provide the $LSJ$ labeling system used in databases such as the NIST ASD ~\citep{Kramida.2015.V.p} and CHIANTI~\citep{DelZanna.2015.V582.p56,Dere.1997.V125.p149}, the wave functions are
transformed  from a \emph{jj}-coupled CSF basis into a
$LSJ$-coupled CSF basis using the methods developed by Gaigalas \citep{Gaigalas.2004.V157.p239,Gaigalas.2017.V5.p6}.

\section{Evaluation of data}\label{sect:com}

\subsection{Energy Levels and Lifetimes}\label{sect:en}
The excitation energies and lifetimes for the 143 lowest states of Fe~XII from the largest CI calculation with orbital sets with 
principal quantum numbers up to $n = 8$ and with orbital angular momenta up to $l = 6$ are displayed in Table \ref{tab.lev.Fe}. For comparison, compiled 
excitation energies from NIST ASD ~\citep{Kramida.2015.V.p} and compiled and calculated energies from CHIANTI version 8~\citep{DelZanna.2015.V582.p56,Dere.1997.V125.p149} 
are also given along with  
excitation energies from the MR-MP calculations by~\cite{Vilkas}, MCHF-BP with non-orthogonal orbitals by~\cite{Tayal.2011.V97.p481} and super-structure (SS) calculations by~\cite{SS2005}. Lifetimes from CHIANTI as well as from the three latter calculations are also given. 
We note that the CHIANTI v.8 data were obtained by  \cite{DZFEXII} with semi-empirical corrections.
The agreement between the present MCDHF/CI calculations and observations is, although not as excellent as for the MR-MP calculation by Vilkas, very good. Also the calculations by Tayal reproduce the excitation energies to a high degree. The SS calculations, which mainly aim at providing target states for scattering calculations, are semi-empirically scaled. For the states where there are no experimental data, 
the SS calculations differ strongly from the more accurate MCDHF/CI and MR-MP calculations. 

For P-like ions the ground state is a quartet state, and it is known that high spin states
converge more rapidly with respect to the increasing orbital space than the other states \citep{2005JChPh.123c4302G,Jonsson.2017.V5.p16}. As a consequence there is a energy shift in the 
MCDHF/CI 
calculations so that the excited states are somewhat too high relative to the ground state. To quantify the shift we computed the mean level deviation, $MLD$, according to    
\begin{equation}
\label{eq:MLD}
MLD = \frac{1}{N} \sum_{i = 1}^N |E_{obs}(i) - E_{cal}(i) + ES |,
\end{equation}
where $N$ is the number of states. The energy shift, $ES$, is chosen as to minimize the sum. For MCDHF/CI we have a shift $ES = 367$ cm$^{-1}$ and $MLD = 213$ cm$^{-1}$. The corresponding 
values for MR-MP and the MCHF-BP calculations are $ES = 233$ cm$^{-1}$ and $MLD = 271$ cm$^{-1}$ and $ES = 500$ cm$^{-1}$ and $MLD = 3993$ cm$^{-1}$, respectively. 
Thus the spread for the MCDHF/CI calculation is small. By subtracting $ES$ from the energies of the 
excited states we greatly improve the predictive power of the calculated values, facilitating the use of the transition energies for identifying lines in observed spectra.  The mean  level deviation $ES$ analysis is summarized in Table \ref{tab.MLD.all}.

Turning to the lifetimes we compare our lifetimes with theoretical lifetimes in the CHIANTI database \citep{DelZanna.2015.V582.p56,Dere.1997.V125.p149} and lifetimes from the calculations by \cite{Tayal.2011.V97.p481} and \cite{Vilkas}.
The latter two
calculations only include the E1 transitions, and thus there are no lifetimes for the states of the ground term, nor from states 18 and 25. The MCDHF/CI lifetimes are computed in both the length form and velocity form. Excluding the states of the ground term configuration and 18 and 25, which are governed by higher multipoles, the lifetimes in the two forms agrees to within 0.3~\% in mean. This agreement is highly satisfactory and indicate accurate values \citep{atoms2020215}. The consistency of the calculated lifetimes 
from the different methods is fair, but with clear anomalies for some states; for state 23 the lifetimes varies 
between \mbox{$1.0 \times 10^{-4}$ s} and \mbox{$5.4 \times 10^{-6}$ s}. Comparing with the experimental lifetimes of \cite{LIFE} we see that the calculated lifetimes for state 19 all gather around 
 \mbox{$3.0 \times 10^{-8}$ s}, which is about a factor of four  longer than the experimental value \mbox{$7.3 \pm 1\times 10^{-9}$ s}. Large differences between calculated and experimental lifetimes are also found for states in Co XIII and Cu XV. New experiments are needed to clarify these differences.

In  Table \ref{tab.lev.all} (the full
table is available on-line) we give calculated excitations energies as well as energies shifted according to the $ES$ values from a mean level deviation analysis that is summarized in Table \ref{tab.MLD.all}. For comparison, excitation energies from \cite{REFDATA}; Cr X, Mn XI, Fe XII, Co XIII, Ni XIV, \cite{REFDATACU}; Cu XV, \cite{REFDATAZN}; Zn XVI, compiled and reported in NIST ASD, are also given as well as excitation energies from the MR-MP calculations by \cite{Vilkas}.
We can see excellent agreement in all the cases where an experimental energy is available (except for Fe XII, see below). 
The excitation energies from the shifted energies are estimated to be accurate to within 0.05~\% and are 
therefore helpful in identifying transitions in spectra from laboratory and astrophysical spectra.
 Finally,  the table gives the
calculated lifetimes in the length and velocity gauges, where the former values are generally believed to be more accurate.

\subsection{Fe XII line identifications}

Table~\ref{tab.rcichi.all} lists 
our  shifted energies for the 41 lowest  Fe XII  levels ($3s^23p^3$, $3s3p^4$, $3s^23p^23d$),
 compared with three sets of  experimental data: those in the NIST database, those suggested by \cite{DZ05} and those
available in  the  CHIANTI database version 8 \citep{DelZanna.2015.V582.p56}.
The  NIST energies are mainly originating from laboratory measurements of B.C. Fawcett and 
collaborators in the 1970s (see, e.g. \citealt{bromage_etal:78}). \cite{DZ05} used available astrophysical
spectra, in addition to Fawcett's  laboratory plates and semi-empirically adjusted theoretical energies, 
to suggest new identifications of almost all the previously-unkown  $3s^23p^23d$ levels.  
Aside from two exceptions, we can see from Table~\ref{tab.rcichi.all} an excellent agreement
between the present energies  and
those listed by \cite{DZ05}, in terms of both the
suggested experimental energies and the predicted ones,
for the 3s$^2$ 3p$^2$ 3d levels.

Later, \cite{DZHINODE} presented a table of wavelengths
and calibrated radiances of coronal lines observed off the solar limb with  Hinode EIS.  On the basis of these observations, two new Fe~XII identifications were suggested, and later introduced in  CHIANTI  version 8. 
The first one is the main decay from the $3s^23p^23d~^4D_{3/2}$ (level 21).
It was tentatively assigned to a self-blend at 249.38~\AA,
changing the predicted energy from 448071 cm$^{-1}$ (which is very close to our
predicted one) to 447076 cm$^{-1}$. Our energy predicts that the line should be
at 248.56~\AA. There is indeed a weak line in the coronal EIS spectrum observed at 248.50~\AA. The observed intensity, relative to the other Fe XII known lines, is in broad agreement with the theoretical ratio, according to CHIANTI version 8. This weak line would normally be blended with an O V transition in on-disk observations. However, the observations of \cite{DZHINODE} were off-limb where no emission in low-temperature lines such O V was present. Therefore, it is likely that this is the correct assignment.

The second tentative identification from  \cite{DZHINODE} regards the
main decay from the $3s^23p^23d~^4D_{5/2}$ (level 22), with a
weak coronal line observed at 245.89~\AA. In this case the
experimental energy is almost the same (452775 cm$^{-1}$)
 as our predicted one (452671 cm$^{-1}$), so it appears that this tentative
identification is correct.

Finally, the main discrepancy with the  \cite{DZ05}
identifications regards the  $3s^23p^23d~^2S_{1/2}$ (level 38),
which has a large difference (3674 cm$^{-1}$) with our predicted value.
Its main decay, to  the $3s^2 3p^3~^2P_{3/2}$ level, is a
transition which becomes strong at high densities.
On the basis of their  estimated energy for this level,
\cite{DZ05}  assigned this decay to a previously
unidentified line present in the laboratory plates at  201.76~\AA.
On the other hand, the original NIST energy was 579630 cm$^{-1}$,
i.e. very close to our predicted energy (579827 cm$^{-1}$)
and to the value (579853 cm$^{-1}$) from the MR-MP calculations by \cite{Vilkas}.
The NIST energy originated from the identification, on the same
laboratory plates, with the 200.356~\AA\ line by \cite{bromage_etal:78}.

It is clear that the original identification by  Bromage et al.
was correct. Indeed, this issue was pointed out by
\cite{Beiersdorfer.2014.V788.p25}.
Excellent agreement was found between the
EBIT spectrum and the CHIANTI atomic data, which included the
 \cite{DZ05}  identifications for P-like Fe and
similar ones obtained by Del Zanna on a number of other coronal iron ions.
Only one main exception stood out, the line at 200.356~\AA,
which is strong in the EBIT spectra.

Regarding the $3s 3p^3 3d$ levels, we note that none of them
were previously identified with certainty. Lines within the
$3s 3p^4$-$3s 3p^3 3d$ transition array are in fact much weaker
compared to the decays from the $3s^2 3p^2 3d$ levels.
The strongest decay, the dipole-allowed
$^4P_{5/2}$--$^4D_{7/2}$ line (transition 6--84)
was tentatively identified by  \cite{DZ05}   with the
191.045~\AA\  line, although later  estimates of the
energies  by \cite{DZFEXII}  put this identification in doubt.
The present {\em ab initio} energy for the  upper level is 810021 cm$^{-1}$.
Applying the same ES shift of Table~\ref{tab.MLD.all} (367 cm$^{-1}$), we
predict the decay to be at 186.8~\AA. Considering the intensity
of this decay and the Hinode EIS observations \citep{DZHINODE}, 
there are only two  possibilities  close in wavelength:
the first is the 186.88~\AA\ line, which is already a known blend of
three transitions (two already from Fe XII); the second is the 187.00~\AA\
line. 
It is very much likely that the second possibility is the correct
identification, given that the 187.00~\AA\  line is currently
unidentified and has an observed intensity close to the predicted one (according to CHIANTI version 8), relative to the intensities of the known Fe XII lines. 
The energy of the upper level would then be only
522 cm$^{-1}$ lower than our prediction.
Moreover, the second strongest transition from these
levels is the decay from the close $3s 3p^3 3d~^4D_{5/2}$ (level 87)
to the $3s 3p^4~^4P_{3/2}$ (level 7).
Applying the same shift of 522 cm$^{-1}$ to this $^4D_{5/2}$ upper level,
we find a wavelength of 189.08~\AA, very close to an observed coronal line
at 189.12~\AA,  identified by \cite{DZHINODE} 
as an Fe XI transition which was significantly (50~\%)  blended.
The predicted intensity of the Fe XII transition
would explain the blend.
If we assumed that the 6--84 transition would blend the 186.88~\AA\ line,
we would expect the 7--87 transition to fall at 189.0~\AA, where there is
a coronal line, but its intensity is entirely due to Fe~XI, as shown
in  \cite{DZHINODE}.
Finally, a weaker decay of level 87 to level 6 ($3s 3p^4 ~ ^4P_{5/2}$)
should also be observable by Hinode EIS at 185.70~\AA, assuming
the second option (with a shift of 522 cm$^{-1}$). Indeed there is a
weak coronal unidentified line  at 185.68~\AA.

Another $3s 3p^3 3d$ level which should produce a weak but
observable line is the lower  $^4D_{7/2}$ (level 50), with a
decay to the  $3s^2 3p^2 3d  ~^4P_{5/2}$ (level 27).
Its predicted intensity, using the \cite{DZFEXII}
atomic data, is about 1/10 the intensity of the decay of the
$3s^2 3p^2 3d ~^4F_{9/2}$  (the 6--18 transition),
identified by \cite{DZ05}   with the  592.6~\AA\ line.
Our predicted energy, with the 367 cm$^{-1}$ adjustment, predicts
this decay to be at 592.7~\AA.
Given that there are no other obvious coronal lines nearby, and the
fact that the intensity of the  6--18 transition was only able to
explain about 70~\% of the observed intensity  \citep{DZ05},
plus the fact that this line is broader than  the other nearby coronal lines,
we conclude that it is very likely that the 27--50 transition is
blending the  592.6~\AA\ line.

\subsection{Transition rates}\label{sect:trans}
\begin{figure*}[h]
	\epsscale{1.05}
	\plotone{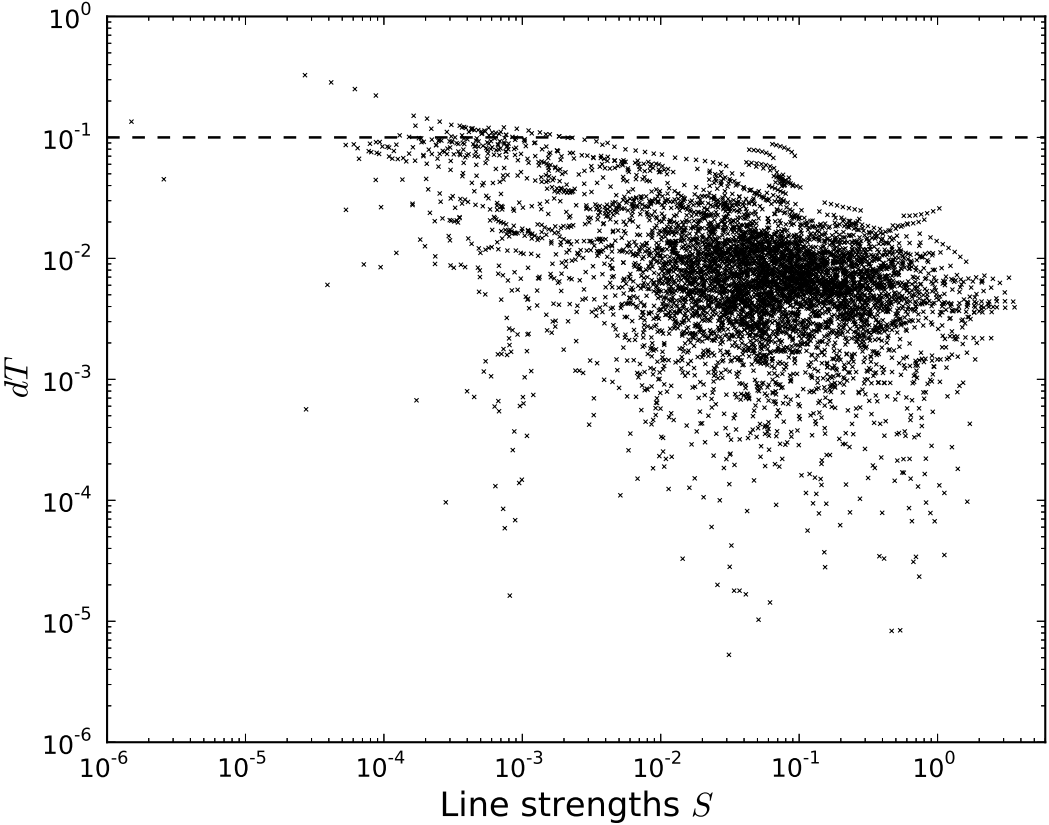}
	\caption{Scatter plot of $dT$, the relative difference between the transition rates in length and velocity form, versus the line strength $S$ for transitions in Fe XII  with branching fractions BF $> 1~\%$. $dT$ is well below 5~\% for the majority of the stronger transitions. \label{fig.trans.all}}
\end{figure*}

In Table \ref{tab.trans.all} (the full
table is available on-line), wavelengths, transition
rates $A_{ji}$, weighted oscillator strengths $gf_{ji}$ and line strength $S_{ji}$, the latter in the length form, are given along with branching fractions ($BF_{ji} = \frac{A_{ji}}{\sum \limits_{k=1}^{j-1} A_{jk}}$ ) and the uncertainty indicator $dT$. 
For most of the stronger E1 transitions $dT$ is well below \mbox{5~\%}. For the weaker transitions, as
displayed in the scatterplot  of $dT$ versus the line strength $S$ for transitions in
Fe XII  with branching fractions \mbox{BF $> 1~\%$}, the uncertainty $dT$ is somewhat larger, from 5~\%
percent up to 30~\% (Figure \ref{fig.trans.all}). The weaker E1 transitions are often intercombination
transitions, where the smallness of the rates comes
from cancellations in the contributions to the transition matrix
elements, compare \cite{PhysRevA.51.2020}, or two-electron-one-photon transitions, which have  zero rate 
in the lowest approximation and where the transition is opened as corrective basis functions are included in the wave function
expansions \citep{0953-4075-43-3-035005}. These two types of transitions are still challenging to theory, requiring very large CSF expansions. 

In Table \ref{tab.trans.DZ}, our calculate transition rates and wavelengths for the brightest coronal lines in Fe XII  are compared with the
values given by \cite{DZ05} and NIST. There is a good agreement between the current transition rates and the rates given by \cite{DZ05}.
There are however a few noticeable differences:  for line 3 - 23 the rate from the current calculation is almost three times larger than the one from \cite{DZ05},
for 2 -- 30 the value from the current calculations agrees reasonably well with the value from \cite{DZ05}, both being more than two times larger than the value given by NIST.
Looking at the wavelengths we see that the current calculation provide excellent predictions. 
For the last transition (6 -- 84) there is a noticeable difference with the wavelength
by \cite{DZ05}, as we have previously discussed.

\section{Summary and conclusions}
We have performed MCDHF and subsequent CI calculations including Breit and QED effects for
states of the $3s^23p^3$, $3s3p^4$, $3s^23p^23d$, $3s3p^33d$, $3p^5$, $3s^23p3d^2$ configurations in
P-like ions from Cr X to Zn XVI. Valence-valence and core-valence electron correlation effects are included in large basis
expansions. Excitation energies, lifetimes, and transition rates are given.
Energies from the CI calculations are in overall excellent agreement
with the few observations available for  the $3s^23p^3$, $3s3p^4$, $3s^23p^23d$ levels.
It is clear that the computed wavelengths are accurate enough to directly aid  line identification in the spectra.
In particular, we have reassessed the previous identifications of the important Fe XII $3s^23p^23d$ levels,
confirming most of the previous suggestions. We have also suggested new identifications of 
a few  $3s3p^33d$ levels, on the basis of our calculated energies, Hinode EIS spectra 
and the \cite{DZFEXII} atomic data.
Further work on the other ions in the sequence is in progress, to see if 
more spectral lines can be identified.

Uncertainties of the transition rates are estimated
 by the relative difference $dT$ between the values in length 
and velocity form. 
For most of the stronger transitions
$dT$ is well below 5~\%. For the weaker transitions the uncertainty
$dT$ is somewhat larger, from 5~\%  up to 30~\%. We are therefore confident 
that the transition rates are highly accurate. Data from the present study will serve
as a benchmark for further calculations.

\acknowledgments
We acknowledge the support from the National Key Research and Development Program of China under Grant No. 2017YFA0402300, the National Natural Science Foundation of China (Grant Grant No. 11703004 and No. 11674066) and the Nature Science Foundation of Hebei Province, China (A2017201165). This work is also supported by the Swedish research council under contract 2015-04842. K.W. expresses his gratitude for the support from the visiting researcher program at the Fudan University. G. D. Z. acknowledges the support from STFC (UK) through the consolidated grant RG 84192. 

\software{GRASP2K \citep{Jonsson.2007.V177.p597,Jonsson.2013.V184.p2197}} is used in the present work.

\clearpage
\bibliographystyle{aasjournal}
\bibliography{ref}


\clearpage
\listofchanges

\clearpage

\clearpage

\begin{deluxetable*}{clrrrrrrrcccc}
	\tabletypesize{\tiny}
	\setlength{\tabcolsep}{4pt}
	\tablecaption{Energy levels (in cm$^{-1}$) relative to the ground state along with radiative lifetime (in s) for the lowest 143 states of \ion{Fe}{12}. 
	$E_{\rm CI}$-the present calculations; $E_{\rm NIST}$ observed energies listed in the NIST database \citet{Kramida.2015.V.p}; $E_{\rm{CHI1}}$ and $E_{\rm{CHI2}}$ observed and calculated energies, respectively, listed in the CHIANTI database 
	\citet{Dere.1997.V125.p149,DelZanna.2015.V582.p56}; 
$E_{\rm{MRMP}}$ MR-MP calculations by \citet{Vilkas}; $E_{\rm{MCHF-BP}}$ MCHF-BP calculations by \citet{Tayal.2011.V97.p481}; $E_{\rm{SS}}$ SS calculations by~\citet{DZ05};
	$\tau_{\rm{CI}}$ lifetimes from the present calculations;  $\tau_{\rm{CHI2}}$ calculated lifetimes given in the CHIANTI database; 
	$\tau_{\rm{MCHF-BP}}$ MCHF-BP calculations by Tayal; $\tau_{\rm{MRMP}}$ MR-MP calculations by Vilkas. 
\label{tab.lev.Fe}
}
	\tablehead{\colhead{Key} & \colhead{Level} & \colhead{$E_{\rm CI}$} & \colhead{$E_{\rm NIST}$} & \colhead{$E_{\rm CHI1}$} & \colhead{$E_{\rm{CHI2}}$} & \colhead{$E_{\rm{MRMP}}$} & \colhead{$E_{\rm{MCHF-BP}}$} &  \colhead{$E_{\rm{SS}}$} &  \colhead{$\tau_{\rm{CI}}$} & \colhead{$\tau_{\rm{CHI2}}$} & \colhead{$\tau_{\rm{MCHF-BP}}$} 
	& \colhead{$\tau_{\rm{MRMP}}$} }
	\startdata
1  &$3s^{2}\,3p^{3}(^{4}S)~^{4}S_{3/2}^{\circ}$                       &       0  &       0  &       0  &       0  &       0  &       0  &       0  &  \nodata    &  \nodata    &  \nodata    &  \nodata    \\ 
2  &$3s^{2}\,3p^{3}(^{2}D)~^{2}D_{3/2}^{\circ}$                       &   41771  &   41566  &   41555  &   41556  &   41518  &   41567  &   41555  &  1.845E-02  &  2.257E-02  &  \nodata    &  \nodata    \\ 
3  &$3s^{2}\,3p^{3}(^{2}D)~^{2}D_{5/2}^{\circ}$                       &   46319  &   46075  &   46088  &   46088  &   46049  &   45967  &   46088  &  3.259E-01  &  3.808E-01  &  \nodata    &  \nodata    \\ 
4  &$3s^{2}\,3p^{3}(^{2}P)~^{2}P_{1/2}^{\circ}$                       &   74372  &   74109  &   74107  &   74107  &   74137  &   74060  &   74107  &  3.845E-03  &  4.207E-03  &  \nodata    &  \nodata    \\ 
5  &$3s^{2}\,3p^{3}(^{2}P)~^{2}P_{3/2}^{\circ}$                       &   80762  &   80515  &   80515  &   80515  &   80536  &   80481  &   80515  &  1.599E-03  &  1.688E-03  &  \nodata    &  \nodata    \\ 
6  &$3s~^{2}S\,3p^{4}(^{3}P)~^{4}P_{5/2}$                             &  274498  &  274373  &  274373  &  274373  &  274279  &  274930  &  274373  &  6.118E-10  &  6.041E-10  &  6.210E-10  &  6.750E-10  \\ 
7  &$3s~^{2}S\,3p^{4}(^{3}P)~^{4}P_{3/2}$                             &  284126  &  284005  &  284005  &  277374  &  283908  &  283863  &  284005  &  5.659E-10  &  6.049E-10  &  5.810E-10  &  6.250E-10  \\ 
8  &$3s~^{2}S\,3p^{4}(^{3}P)~^{4}P_{1/2}$                             &  288451  &  288307  &  288307  &  288307  &  288213  &  288000  &  288307  &  5.355E-10  &  5.332E-10  &  5.530E-10  &  5.920E-10  \\
\enddata
\end{deluxetable*}

\clearpage

\begin{table}
\caption{Mean level deviation $MLD$ and energy shift $ES$ (in cm$^{-1}$) for the MCDHF/CI and MRMP calculations obtained from Eq.~\ref{eq:MLD}. \label{tab.MLD.all}}
\begin{tabular}{rrrrr} \hline
    & \multicolumn{2}{c}{MCDHF/CI} & \multicolumn{2}{c}{MR-MP} \\
$Z$ & $ES$    &   $MLD$     &    $ES$    &   $MLD$    \\   \hline
24  & 333    &   151     & \nodata &\nodata\\
25  & 383    &   144     & \nodata& \nodata\\
26  & 367    &   213     &    233    &   271    \\
27  & 326    &   139     &     47    &   216    \\
28  & 345    &   122     &     68    &   182    \\
29  & 385    &   206     &     72    &   232    \\
30  & 280    &   146     &     66    &   202    \\ \hline
\end{tabular}
\end{table} 

\begin{deluxetable*}{crlrrrrrc}
	\tabletypesize{\tiny}
	\setlength{\tabcolsep}{4pt}
	\tablecaption{
Energy levels (in cm$^{-1}$) relative to the ground state along with radiative lifetime (in s) for the lowest 143 states of  Cr~X, Mn~XI, Fe~XII, Co~XIII, Ni~XIV, Cu~XV, Zn~XVI. 
	$E_{\rm CI1}$ the present {\em ab initio} calculations; $E_{\rm CI2}$ the present calculations where energies of excited states have been shifted by a constant $ES$ given in Table~\ref{tab.MLD.all} (see text for more details). $E_{\rm NIST}$ observed energies listed in the NIST database \citet{Kramida.2015.V.p}; $E_{\rm MRMP}$ MR-MP calculations by \citet{Vilkas}; 
	$\tau_{\rm{l}}$ lifetimes in s from the present calculations in length gauge;  $\tau_{\rm{v}}$ lifetimes from the present calculations in velocity gauge. \label{tab.lev.all}}
	\tablehead{\colhead{$Z$} & \colhead{Key} & \colhead{Level}  & \colhead{$E_{\rm CI1}$} & \colhead{$E_{\rm CI2}$} & \colhead{$E_{\rm NIST}$} & 
	\colhead{$E_{\rm MRMP}$} & \colhead{$\tau_{\rm l}$} & \colhead{$\tau_{\rm v}$}  }
	\startdata
28 &    1 &   $3s^{2}\,3p^{3}(^{4}S)~^{4}S_{3/2}^{\circ}$                    &        0 &       0 &       0  &   0        &    \nodata   &   \nodata   \\ 
28 &    2 &   $3s^{2}\,3p^{3}(^{2}D)~^{2}D_{3/2}^{\circ}$                    &    45978 &   45633 &   45767  &     45729  &   5.502E-03  &  \nodata    \\ 
28 &    3 &   $3s^{2}\,3p^{3}(^{2}D)~^{2}D_{5/2}^{\circ}$                    &    53798 &   53453 &   53569  &     53534  &   7.165E-02  &  \nodata    \\ 
28 &    4 &   $3s^{2}\,3p^{3}(^{2}P)~^{2}P_{1/2}^{\circ}$                    &    85396 &   85051 &   85126  &     85151  &   1.513E-03  &  \nodata    \\ 
28 &    5 &   $3s^{2}\,3p^{3}(^{2}P)~^{2}P_{3/2}^{\circ}$                    &    96909 &   96564 &   96630  &     96678  &   6.313E-04  &  \nodata    \\ 
28 &    6 &   $3s~^{2}S\,3p^{4}(^{3}P)~^{4}P_{5/2}$                          &   316478 &  316133 &  316343  &    316237  &   4.768E-10  &  4.631E-10  \\ 
28 &    7 &   $3s~^{2}S\,3p^{4}(^{3}P)~^{4}P_{3/2}$                          &   330975 &  330630 &  330837  &    330739  &   4.291E-10  &  4.208E-10  \\ 
28 &    8 &   $3s~^{2}S\,3p^{4}(^{3}P)~^{4}P_{1/2}$                          &   336908 &  336563 & \nodata  &    336644  &   3.988E-10  &  3.924E-10  \\
\enddata
\tablecomments{ Table~\ref{tab.lev.all} is published in its entirety in the machine-readable format. The values for $Z=28$ are shown here for guidance regarding its form and content.}
\end{deluxetable*}

\clearpage

\begin{deluxetable*}{rlrrrrr}
	\tabletypesize{\tiny}
	\setlength{\tabcolsep}{4pt}
	\tablecaption{Comparison of excitation energies (in cm$^{-1}$) for the lowest 41 levels Fe XII. 
	$E_{\rm CI2}$ the present calculations where energies of excited states have been shifted by a constant $ES$ given in Table \ref{tab.MLD.all} 
	(see text for more details). $E_{\rm NIST}$ are the NIST energies.
 $E_{\rm DM}$ lists the observed energies recommended by \cite{DZ05} 
(values in brackets indicate estimated energies), while 
 $E_{\rm CHI}$ lists the observed energies as  in the 
   CHIANTI database version 8~\citep{DelZanna.2015.V582.p56,Dere.1997.V125.p149}. The superscript $^o$ indicates
energies that appear to be incorrect. 
	$\Delta E = E_{\rm CI2} - E_{\rm CHI}$.\label{tab.rcichi.all} }
	\tablehead{\colhead{Key} & \colhead{Level}  & \colhead{$E_{\rm CI2}$} & 
              \colhead{$E_{\rm NIST}$}  &     \colhead{$E_{\rm DM}$}  & \colhead{$E_{\rm CHI}$} & \colhead{$\Delta E$} }
	\startdata
 1 &  $3s^{2}\,3p^{3}(^{4}S)~^{4}S_{3/2}^{\circ}$                   &       0 &      0 &      0  &       0  &        0  \\  
2  &  $3s^{2}\,3p^{3}(^{2}D)~^{2}D_{3/2}^{\circ}$                   &   41404 &  41566 &  41555  &   41555  &    -151  \\
3  &  $3s^{2}\,3p^{3}(^{2}D)~^{2}D_{5/2}^{\circ}$                   &   45952 &  46075 &  46088  &   46088  &    -136  \\
4  &  $3s^{2}\,3p^{3}(^{2}P)~^{2}P_{1/2}^{\circ}$                   &   74005 &  74372 &  74107  &   74107  &    -102  \\
5  &  $3s^{2}\,3p^{3}(^{2}P)~^{2}P_{3/2}^{\circ}$                   &   80395 &  80515 &  80515  &   80515  &    -120  \\
6  &  $3s~^{2}S\,3p^{4}(^{3}P)~^{4}P_{5/2}$                         &  274131 & 274373 & 274373  &  274373  &    -242  \\
7  &  $3s~^{2}S\,3p^{4}(^{3}P)~^{4}P_{3/2}$                         &  283759 & 284005 & 284005  &  284005  &    -246  \\
8  &  $3s~^{2}S\,3p^{4}(^{3}P)~^{4}P_{1/2}$                         &  288084 & 288307 & 288307  &  288307  &    -223  \\
\enddata                                                                          
\end{deluxetable*}                                                                

\newpage

\begin{deluxetable*}{llllllllll}
	\tabletypesize{\tiny}
	\setlength{\tabcolsep}{4pt}
	\tablecaption{Transition wavelengths $\lambda$ (in {\AA}), transition rates $A$ (in s$^{-1}$), weighted oscillator strengths $gf$ and line strengths $S$ (in a.u.) 
	between the lowest 143 states of  Cr X, Mn XI, Fe XII, Co XIII, Ni XIV, Cu XV, Zn XVI. All transition data are in length form. Type is the type of the multipole, BF is the branching fraction from the upper level and
		$dT$ is the relative difference of the transition rates in length and velocity form as given by Eq.~(3). Only transitions with BF  $\geq 10^{-5}$ are presented. \label{tab.trans.all} }
	\tablehead{\colhead{$Z$} & \colhead{$i$} & \colhead{$j$} &    \colhead{Type}  & 
	\colhead{$\lambda$} & \colhead{$A$} & \colhead{$gf$} & \colhead{$S$} & \colhead{BF} & \colhead{$dT$}      }
	\startdata
24 &  1&   2& M1& 2.6802E+03& 1.298E+01& 5.591E-08& 3.705E-02& 9.984E-01&          \\
24 &  1&   2& E2& 2.6802E+03& 2.144E-02& 9.238E-11& 1.059E-02& 1.650E-03& 1.222E-03\\
24 &  1&   3& M1& 2.5205E+03& 3.874E-01& 2.214E-09& 1.380E-03& 6.900E-01&          \\
24 &  1&   3& E2& 2.5205E+03& 4.305E-02& 2.460E-10& 2.346E-02& 7.667E-02& 1.019E-03\\
24 &  2&   3& M1& 4.2302E+04& 1.310E-01& 2.109E-07& 2.206E+00& 2.333E-01&          \\
24 &  1&   4& M1& 1.5579E+03& 6.488E+01& 4.722E-08& 1.819E-02& 6.999E-01&          \\
24 &  1&   4& E2& 1.5579E+03& 2.118E-02& 1.541E-11& 3.471E-04& 2.285E-04& 9.716E-04\\
24 &  2&   4& M1& 3.7203E+03& 2.703E+01& 1.122E-07& 1.032E-01& 2.915E-01&          \\
\enddata
	\tablecomments{ Table~\ref{tab.trans.all} is published in its entirety in the machine-readable format. Part of the values for $Z=24$ are shown here for guidance regarding its form and content.}
\end{deluxetable*}

\begin{deluxetable*}{llllllllll}
	\tabletypesize{\tiny}
	\setlength{\tabcolsep}{4pt}                                                         
	\tablecaption{Transition wavelengths $\lambda$ (in {\AA}) and transition rates $A$ (in s$^{-1}$) for the brightest lines in Fe XII. Column 6 and 7 transition wavelengths and rates from this work, column 8 and 9  transition wavelengths and rates from \cite{DZ05}. $\lambda_{CI}$ and $A_{CI}$ are from this work, 
		$\lambda_{DZ}$ and $A_{DZ}$ from \cite{DZ05}.
	Column 10 transition rates from NIST. This table is an adaptation of Table 3 from \cite{DZ05}.
\label{tab.trans.DZ} }
	\tablehead{\colhead{$i$} & \colhead{$j$} & \colhead{lower level} &  \colhead{upper level}  & \colhead{Type}  & 
	\colhead{$\lambda_{CI}$} & \colhead{$A_{CI}$} & \colhead{$\lambda_{DZ}$} & \colhead{$A_{DZ}$} & \colhead{$A_{NIST}$}        }
	\startdata
1  &    2 &  $3s^{2}\,3p^{3}(^{4}S)~^{4}S_{3/2}^{\circ}$               &  $3s^{2}\,3p^{3}(^{2}D)~^{2}D_{3/2}^{\circ}$              &  M1  &   2394.0  &   5.415E+01 &   2406.41  $\pm$  0.06  &  5.6E+01 &  4.8E+01   \\ 
1  &    3 &  $3s^{2}\,3p^{3}(^{4}S)~^{4}S_{3/2}^{\circ}$               &  $3s^{2}\,3p^{3}(^{2}D)~^{2}D_{5/2}^{\circ}$              &  M1  &   2158.9  &   2.071E+00 &   2169.76  $\pm$  0.05  &  2.3E+00 &  1.7E+00   \\ 
1  &    5 &  $3s^{2}\,3p^{3}(^{4}S)~^{4}S_{3/2}^{\circ}$               &  $3s^{2}\,3p^{3}(^{2}P)~^{2}P_{3/2}^{\circ}$              &  M1  &   1238.2  &   3.447E+02 &   1242.01  $\pm$  0.01  &  3.5E+02 &  3.2E+02   \\ 
1  &    4 &  $3s^{2}\,3p^{3}(^{4}S)~^{4}S_{3/2}^{\circ}$               &  $3s^{2}\,3p^{3}(^{2}P)~^{2}P_{1/2}^{\circ}$              &  M1  &   1344.6  &   1.888E+02 &   1349.40  $\pm$  0.02  &  1.9E+02 &  1.8E+02   \\ 
2  &    5 &  $3s^{2}\,3p^{3}(^{2}D)~^{2}D_{3/2}^{\circ}$               &  $3s^{2}\,3p^{3}(^{2}P)~^{2}P_{3/2}^{\circ}$              &  M1  &   2564.7  &   1.978E+02 &   2566.77  $\pm$  0.13  &  2.0E+02 &  2.0E+02   \\ 
3  &    5 &  $3s^{2}\,3p^{3}(^{2}D)~^{2}D_{5/2}^{\circ}$               &  $3s^{2}\,3p^{3}(^{2}P)~^{2}P_{3/2}^{\circ}$              &  M1  &   2903.4  &   7.923E+01 &   2904.70  $\pm$  0.17  &  8.2E+01 &  \nodata   \\ 
2  &    4 &  $3s^{2}\,3p^{3}(^{2}D)~^{2}D_{3/2}^{\circ}$               &  $3s^{2}\,3p^{3}(^{2}P)~^{2}P_{1/2}^{\circ}$              &  M1  &   3067.3  &   7.005E+01 &   3072.06  $\pm$  0.19  &  7.2E+01 &  \nodata   \\ 
18 &   25 &  $3s^{2}\,3p^{2}(^{3}P)~^{3}P\,3d~^{4}F_{9/2}$             &  $3s^{2}\,3p^{2}(^{1}D)~^{1}D\,3d~^{2}G_{9/2}$            &  M1  &   1837.8  &   1.176E+02 &   1847.23  $\pm$  11.19 &  1.2E+02 &  \nodata   \\
\enddata
\end{deluxetable*}

\clearpage

\end{document}